\documentclass[aps,preprint]{revtex4}%
\usepackage{amsfonts}
\usepackage{amsmath}
\usepackage{amssymb}
\usepackage{graphicx}%
\begin{document}
\preprint{ }

\vspace*{1cm}

\begin{center}
{\Large Two-particle scattering on the lattice: \ }

{\Large Phase shifts, spin-orbit coupling, and mixing angles}\vspace*{0.75cm}

{Bu\={g}ra~Borasoy$^{a}$, Evgeny Epelbaum$^{a,b}$, Hermann~Krebs$^{a,b}$,
Dean~Lee$^{c,a}$, Ulf-G.~Mei{\ss }ner$^{a,b}$}\vspace*{0.75cm}

$^{a}$\textit{Helmholtz-Institut f\"{u}r Strahlen- und Kernphysik (Theorie)
Universit\"{a}t Bonn, }\linebreak\textit{Nu\ss allee 14-16, D-53115 Bonn,
Germany }

$^{b}$\textit{Institut f\"{u}r Kernphysik (Theorie), Forschungszentrum
J\"{u}lich, D-52425 J\"{u}lich, Germany }

$^{c}$\textit{Department of Physics, North Carolina State University, Raleigh,
NC 27695, USA}

\vspace*{0.75cm}

{\large Abstract}
\end{center}

We determine two-particle scattering phase shifts and mixing angles for
quantum theories defined with lattice regularization. \ The method is suitable
for any nonrelativistic effective theory of point particles on the lattice.
\ In the center-of-mass frame of the two-particle system we impose a hard
spherical wall at some fixed large radius. \ For channels without partial-wave
mixing the partial-wave phase shifts are determined from the energies of the
nearly-spherical standing waves. \ For channels with partial-wave mixing
further information is extracted by decomposing the standing wave at the wall
boundary into spherical harmonics, and we solve coupled-channels equations to
extract the phase shifts and mixing angles. \ The method is illustrated and
tested by computing phase shifts and mixing angles on the lattice for spin-1/2
particles with an attractive Gaussian potential containing both central and
tensor force parts.

\pagebreak

\section{Introduction}

There have been several recent studies on the subject of lattice simulations
for low-energy nuclear physics using effective interactions
\cite{Muller:1999cp,Abe:2003fz,Chandrasekharan:2003ub,Chandrasekharan:2003wy,Lee:2004si,Lee:2004qd,Hamilton:2004gp,Seki:2005ns,Lee:2005is,Lee:2005it,Borasoy:2005yc,deSoto:2006pe,Borasoy:2006qn,Lee:2007jd}%
. \ Similar lattice effective field theory techniques have been used to study
cold atomic Fermi systems in the limit of short-range interactions and large
scattering length
\cite{Chen:2003vy,Wingate:2005xy,Bulgac:2005a,Lee:2005fk,Burovski:2006a,Burovski:2006b,Lee:2006hr}%
. \ In nearly all cases the connection between lattice interactions and
physical observables is made using some variant of L\"{u}scher's result
\cite{Luscher:1985dn,Luscher:1986pf,Luscher:1991ux} relating the energy levels
of two-body states in a finite-volume cubic box with periodic boundaries to
the infinite-volume scattering matrix.

L\"{u}scher's method has been extended in a number of different ways.
\ Several studies have looked at asymmetric
boxes\ \cite{Li:2003jn,Feng:2004ua}, while another considered small volumes
where the lattice length $L$ is smaller than the scattering length
\cite{Beane:2003da}. \ There have also been studies of moving frames
\cite{Rummukainen:1995vs, Kim:2005gf}, winding of the interaction around the
periodic boundary \cite{Sato:2007ms}, modifications at finite lattice spacing
\cite{Seki:2005ns}, and techniques to distinguish shallow bound states from
scattering states using Levinson's theorem \cite{Sasaki:2006jn}. \ A very
recent study derived finite volume formulas for systems of $n$ bosons with
short-range repulsive interactions \cite{Beane:2007qr}.

While L\"{u}scher's method has been very useful at low momenta, there is
currently no technique which is able to determine phase shifts on the lattice
at higher energies and higher orbital angular momenta. \ There is also no
technique which can accurately measure spin-orbit coupling and partial-wave
mixing on the lattice. \ The physics of spin-orbit coupling and partial-wave
mixing is difficult to extract using L\"{u}scher's method due to artifacts
generated by the periodic cubic boundary. \ An example of this problem is
shown in the summary and discussion section of this paper. \ Any method
overcoming these theoretical problems would probably not be numerically
practical when applied to lattice simulations of quantum chromodynamics.
\ This is because each individual hadron must be constructed as a bound state
of quark and gluon fields, and the Monte Carlo signal for hadron-hadron
scattering states well above threshold would be very weak. \ However for an
effective lattice theory of fundamental point particles, such a technique
should be numerically viable since the full two-particle spectrum is
relatively easy to compute.

In this paper we discuss a simple method which directly measures two-particle
phase shifts and mixing angles on the lattice. \ In the center-of-mass frame
of the two-particle system we impose a hard spherical wall boundary condition
at some large radius. \ Phase shifts and mixing angles are determined from
properties of the nearly-spherical standing waves produced by the wall
boundary. The method is suitable for any nonrelativistic effective theory of
point particles on the lattice with or without spin. \ We test the method on
the lattice using an attractive Gaussian potential for spin-1/2 particles
containing both central and tensor force parts.

The organization of the paper is as follows. \ We first discuss
representations of the cubic rotational group. \ We then introduce the test
potential and solve the $S=0$ channels and uncoupled $S=1$ channels up to
$J=4$ using the spherical wall method. \ We then discuss coupled partial waves
with a common nodal constraint and the coupled equations needed to solve for
the $L=J-1$ and $J+1$ phase shifts and mixing angle for total angular momentum
$J$. \ Finally we solve the coupled $S=1$ channels on the lattice up to $J=4$
and discuss applications of the technique.

\section{Cubic rotational group}

Our choice of lattice regularization reduces the SO$(3)$ rotational symmetry
of continuous space to the cubic rotational group SO$(3,Z)$. \ This consists
of $24$ group elements generated by products of $\pi/2$ rotations about the
$x$, $y$, $z$\ axes. \ Since SO$(3,Z)$ is discrete, we cannot\ define angular
momentum operators $J_{x}$, $J_{y}$, $J_{z}$ in the usual sense. \ However if
$R_{\hat{z}}\left(  \pi/2\right)  $ is the group element for a $\pi/2$
rotation about the $z$ axis, we can use the SO$(3)$ relation%
\begin{equation}
R_{\hat{z}}\left(  \pi/2\right)  =\exp\left[  -i\frac{\pi}{2}J_{z}\right]
\label{Jz}%
\end{equation}
to define $J_{z}$. \ The eigenvalues of $J_{z}$ are integers modulo 4.
\ $J_{x}$ and $J_{y}$ can be defined in the same manner.

There are five irreducible representations of the cubic rotational group.
\ These are usually written as $A_{1}$, $T_{1}$, $E$, $T_{2}$, and $A_{2}$.
\ Some of their properties and examples in terms of spherical harmonics
$Y_{L,L_{z}}(\theta,\phi)$ are listed in Table \ref{reps}.

\setcounter{table}{0}\begin{table}[tbh]
\caption{Irreducible SO$(3,Z)$ representations}%
\label{reps}
$%
\begin{tabular}
[c]{||c|c|c||}\hline\hline
Re$\text{presentation}$ & $J_{z}$ & Ex$\text{ample}$\\\hline
$A_{1}$ & $0\operatorname{mod}4$ & $Y_{0,0}$\\\hline
$T_{1}$ & $0,1,3\operatorname{mod}4$ & $\left\{  Y_{1,0},Y_{1,1},Y_{1,-1}
\right\}  $\\\hline
$E$ & $0,2\operatorname{mod}4$ & $\left\{  Y_{2,0},\frac{Y_{2,-2}+Y_{2,2}%
}{\sqrt{2}}\right\}  $\\\hline
$T_{2}$ & $1,2,3\operatorname{mod}4$ & $\left\{  Y_{2,1},\frac{Y_{2,-2}%
-Y_{2,2}}{\sqrt{2}},Y_{2,-1}\right\}  $\\\hline
$A_{2}$ & $2\operatorname{mod}4$ & $\frac{Y_{3,2}-Y_{3,-2}}{\sqrt{2}}%
$\\\hline\hline
\end{tabular}
$\end{table}The $2J+1$ elements of the total angular momentum $J$
representation of SO$(3)$ break up into smaller pieces consisting of the five
irreducible representations. \ Examples for $J\leq7$ are shown in Table
\ref{decomp} \cite{Johnson:1982yq}.

\begin{table}[tbh]
\caption{SO$(3,Z)$ decompositions for $J\leq7$}%
\label{decomp}
\begin{tabular}
[c]{||c|c||}\hline\hline
$\text{SO}(3)$ & $\text{SO}(3,Z)$\\\hline
$J=0$ & $A_{1}$\\\hline
$J=1$ & $T_{1}$\\\hline
$J=2$ & $E\oplus T_{2}$\\\hline
$J=3$ & $T_{1}\oplus T_{2}\oplus A_{2}$\\\hline
$J=4$ & $A_{1}\oplus T_{1}\oplus E\oplus T_{2}$\\\hline
$J=5$ & $T_{1}\oplus T_{1}\oplus E\oplus T_{2}$\\\hline
$J=6$ & $A_{1}\oplus T_{1}\oplus E\oplus T_{2}\oplus T_{2}\oplus A_{2}%
$\\\hline
$J=7$ & $T_{1}\oplus T_{1}\oplus E\oplus T_{2}\oplus T_{2}\oplus A_{2}%
$\\\hline\hline
\end{tabular}
\end{table}

In this analysis we consider the scattering of two identical particles and
assume that the interactions are parity conserving. \ Therefore in continuous
space the two particles share the same group representation for intrinsic spin
and the same group representation for all other internal quantum symmetries.
\ As a result each two-particle state must be symmetric or antisymmetric under
parity and symmetric or antisymmetric with respect to internal symmetry
quantum numbers. \ Once these are fixed the overall Fermi or Bose statistics
determines whether the total intrinsic spin combination must be symmetric or antisymmetric.

There are well-known problems associated with massless fermions on the
lattice. \ However for nonrelativistic particles one can easily find a lattice
Hamiltonian or Euclidean action which maintains exact parity invariance and
all internal symmetries.\ \ Therefore the symmetry or antisymmetry of
intrinsic spin on the lattice is unambiguous and the same as in continuous
space. \ For two spin-1/2 particles where the total intrinsic spin is $S=0$ or
$S=1,$ this is enough to specify the intrinsic spin representation completely.
\ If the intrinsic spin is symmetric then the SO$(3,Z)$ representation is
$A_{1}$. \ \ If the intrinsic spin is antisymmetric then the SO$(3,Z)$
representation is $T_{1}$. \ These are in one-to-one correspondence with the
representations $S=0$ and $S=1$ in continuous space, respectively, and so
there is no confusion in borrowing the continuous space names for the two
cases. \ For particles with higher intrinsic spin$,$ there is in general some
unphysical mixing on the lattice among even values of $S$ and among odd values
of $S$. \ The same mixing on the lattice occurs among even values and odd
values of orbital angular momentum $L$ regardless of the intrinsic spin.

\section{Test potential}

There is an endless variety of different interactions one can study on the
lattice. \ For our analysis we choose a simple pedagogical example for which
the continuum limit is relatively easy to compute. \ We choose a bounded
short-range potential producing both a central force and tensor force. \ This
yields all of the essential complications of spin-orbit coupling with
partial-wave mixing while leaving out additional issues of ultraviolet
divergences and singular interactions. \ We take a system of identical
spin-1/2 particles with mass $m$ and a spin-dependent potential,%
\begin{equation}
V(\vec{r}\,)=C\left\{  1+\frac{r^{2}}{R_{0}^{2}}\left[  3\left(  \hat{r}%
\cdot\vec{\sigma}_{1}\right)  \left(  \hat{r}\cdot\vec{\sigma}_{2}\right)
-\vec{\sigma}_{1}\cdot\vec{\sigma}_{2}\right]  \right\}  \exp\left(  -\frac
{1}{2}\frac{r^{2}}{R_{0}^{2}}\right)  .
\end{equation}
The range of the potential is set by the parameter $R_{0}$. \ We leave the
internal symmetry group unspecified and consider all possible combinations of
symmetry or antisymmetry for parity and intrinsic spin.

The tensor operator%
\begin{equation}
S_{12}(\hat{r})=3\left(  \hat{r}\cdot\vec{\sigma}_{1}\right)  \left(  \hat
{r}\cdot\vec{\sigma}_{2}\right)  -\vec{\sigma}_{1}\cdot\vec{\sigma}_{2}
\label{s12}%
\end{equation}
is of considerable general interest in physics. \ It is produced both by
one-pion exchange in low-energy nuclear physics as well as by magnetic and
electric dipole interactions in atoms and molecules. \ The parameters $C$,
$R_{0}$, and $m$ we choose are motivated by low-energy nuclear physics,%
\begin{align}
C  &  =-2\ \text{MeV,}\\
R_{0}  &  =2\times10^{-2}\text{\ MeV}^{-1},\\
m  &  =938.92\text{\ MeV.}%
\end{align}
This produces a very shallow bound state in the $^{3}S(D)_{1}$ channel with
energy $-0.155$ MeV. \ The parenthesis in our spectroscopic notation indicates
a mixture of two partial waves. \ For example $^{3}S(D)_{1}$ indicates a
mixture of $S$ and $D$ partial waves. \ The ordering of the partial waves,
$S(D)$ rather than $D(S),$ indicates that this state becomes a pure $S$ wave
when the tensor part of the interaction is continuously dialed down to zero.
\ This identification is unambiguous in finite volume systems where the energy
levels are discrete.

For total intrinsic spin $S=0$ the tensor operator $S_{12}$ vanishes. \ And so
for this case it suffices to consider just the central part of the potential,%
\begin{equation}
V_{0}(\vec{r}\,)=C\exp\left(  -\frac{1}{2}\frac{r^{2}}{R_{0}^{2}}\right)  ,
\end{equation}
for particles with no intrinsic spin at all. \ We start first with this
simplest case. \ Afterwards we consider $S=1$ for uncoupled channels and then
finally $S=1$ in coupled channels.

\section{Intrinsic spin $S=0$}

\begin{figure}
[ptb]
\begin{center}
\includegraphics[
height=1.9441in,
width=1.9441in
]%
{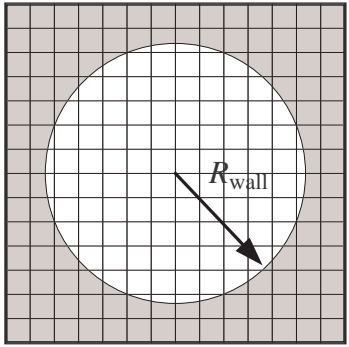}%
\caption{Spherical wall imposed in the center-of-mass frame.}%
\label{spherical_wall}%
\end{center}
\end{figure}

We measure phase shifts by imposing a hard spherical wall boundary on the
relative separation between the two particles at some chosen radius
$R_{\text{wall}}$. \ Viewed in the center-of-mass frame we solve the
Schr\"{o}dinger equation for spherical standing waves which vanish at
$r=R_{\text{wall}}$. \ This is sketched in Fig. \ref{spherical_wall}.%
We are of course assuming that the original lattice system is large enough to
hold a sphere of radius $R_{\text{wall}}$. \ The reason for this spherical
wall is to remove copies of the two-particle interactions due to the periodic
boundaries of the lattice. \ It is this feature that distinguishes our method
from L\"{u}scher's method \cite{Luscher:1991ux}.

For values of $r$ beyond the range of the interaction, the spherical standing
wave in continuous space can be decomposed as a superposition of products of
spherical harmonics and spherical Bessel functions,%
\begin{equation}
\left[  \cos\delta_{L}\times j_{L}(kr)-\sin\delta_{L}\times y_{L}(kr)\right]
Y_{L,L_{z}}(\theta,\phi), \label{wavefunction}%
\end{equation}
where the center-of-mass energy of the spherical wave is%
\begin{equation}
E=2\times\frac{k^{2}}{2m}=\frac{k^{2}}{m},
\end{equation}
and the phase shift for partial wave $L$ is $\delta_{L}$. \ Therefore we know
$k$ from the energy $E$, and the phase shift $\delta_{L}$ is determined by
setting the wavefunction (\ref{wavefunction}) equal to zero at the wall
boundary,%
\begin{equation}
\cos\delta_{L}\times j_{L}(kR_{\text{wall}})=\sin\delta_{L}\times
y_{L}(kR_{\text{wall}}),
\end{equation}%
\begin{equation}
\delta_{L}=\tan^{-1}\left[  \frac{j_{L}(kR_{\text{wall}})}{y_{L}%
(kR_{\text{wall}})}\right]  . \label{simple_phaseshift}%
\end{equation}
On the lattice there is some ambiguity on the precise value of $R_{\text{wall}%
}$ since the components of $\vec{r}$ must be integer multiples of the lattice
spacing. \ We resolve this ambiguity by fine-tuning the value of
$R_{\text{wall}}$ for each standing wave so that $\delta_{L}$ equals zero when
the particles are non-interacting. \ This is illustrated in the following
discussion of the lattice calculation.

\section{Lattice calculation for $S=0$}

Since our method is intended as a tool for simulations of few- and many-body
systems using lattice effective field theory, we present the analysis using
the standard formalism of lattice field theory developed in the literature.
\ For the free part of the lattice action we use the same lattice action
defined in \cite{Borasoy:2006qn} with spatial lattice spacing $a=(100$
MeV$)^{-1}$ and temporal lattice spacing $a_{t}=(70$\ MeV$)^{-1}$. \ We define
$\alpha_{t}$ as the ratio between lattice spacings, $\alpha_{t}=a_{t}/a$.
\ Throughout we use dimensionless parameters and operators, which correspond
with physical values multiplied by the appropriate power of $a$. \ Final
results are presented in physical units with the corresponding unit stated explicitly.

Since the temporal lattice spacing is nonzero we work with transfer matrices
rather than the Hamiltonian directly. \ In simple terms the transfer matrix is
just the exponential of the Hamiltonian $\exp(-H\Delta t)$, where $\Delta t$
equals one temporal lattice spacing. \ More precisely the free-particle
transfer matrix is defined as%
\begin{equation}
M_{\text{free}}\equiv\colon\exp\left(  -H_{\text{free}}\alpha_{t}\right)
\colon,
\end{equation}
where the $::$ symbols indicate normal ordering. \ We use the $O(a^{4}%
)$-improved free lattice Hamiltonian,%
\begin{align}
H_{\text{free}}  &  =\frac{49}{12m}\sum_{\vec{n}}\sum_{j=\uparrow,\downarrow
}a_{j}^{\dagger}(\vec{n})a_{j}(\vec{n})\nonumber\\
&  -\frac{3}{4m}\sum_{\vec{n}}\sum_{j=\uparrow,\downarrow}\sum_{l=1,2,3}%
\left[  a_{j}^{\dagger}(\vec{n})a_{j}(\vec{n}+\hat{l})+a_{j}^{\dagger}(\vec
{n})a_{j}(\vec{n}-\hat{l})\right] \nonumber\\
&  +\frac{3}{40m}\sum_{\vec{n}}\sum_{j=\uparrow,\downarrow}\sum_{l=1,2,3}%
\left[  a_{j}^{\dagger}(\vec{n})a_{j}(\vec{n}+2\hat{l})+a_{j}^{\dagger}%
(\vec{n})a_{j}(\vec{n}-2\hat{l})\right] \nonumber\\
&  -\frac{1}{180m}\sum_{\vec{n}}\sum_{j=\uparrow,\downarrow}\sum
_{l=1,2,3}\left[  a_{j}^{\dagger}(\vec{n})a_{j}(\vec{n}+3\hat{l}%
)+a_{j}^{\dagger}(\vec{n})a_{j}(\vec{n}-3\hat{l})\right]  .
\end{align}
The vector $\vec{n}$ denotes integer-valued coordinate vectors on a spatial
three-dimensional lattice, and $\hat{l}=\hat{1},\hat{2},\hat{3}$ are lattice
unit vectors in each of the spatial directions.

Since the potential energy is finite we can take the lattice potential at
$\vec{n}$ to agree with the continuum potential at $\vec{r}=\vec{n}a$.\ \ For
$S=0$ we can omit the tensor part of the potential, and so the transfer matrix
is%
\begin{equation}
M\equiv\colon\exp\left[  -H_{\text{free}}\alpha_{t}-\frac{\alpha_{t}}{2}%
\sum_{\vec{n}_{1},\vec{n}_{2}}V_{0}(\vec{n}_{1}-\vec{n}_{2})\rho^{a^{\dag}%
,a}(\vec{n}_{1})\rho^{a^{\dag},a}(\vec{n}_{2})\right]  \colon,
\end{equation}
where $\rho^{a^{\dagger},a}(\vec{n})$ is the particle density operator%
\begin{equation}
\rho^{a^{\dagger},a}(\vec{n})=\sum_{j=\uparrow,\downarrow}a_{j}^{\dagger}%
(\vec{n})a_{j}(\vec{n}).
\end{equation}
We calculate eigenvalues of the transfer matrix using the Lanczos method
\cite{Lanczos:1950}. \ In the transfer matrix formalism eigenvalues of the
transfer matrix are interpreted as exponentials of the energy,
\begin{equation}
M\left\vert \Psi\right\rangle =\lambda\left\vert \Psi\right\rangle
=e^{-E\alpha_{t}}\left\vert \Psi\right\rangle .
\end{equation}

\begin{figure}
[ptb]
\begin{center}
\includegraphics[
trim=0.000000in 0.000000in 0.000000in 5.018204in,
height=3.0217in,
width=4.2229in
]%
{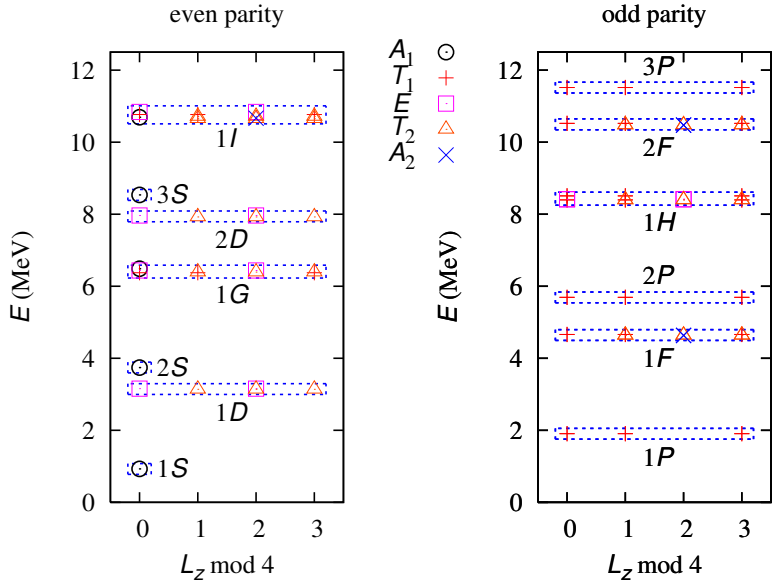}%
\caption{Free particle spectrum for standing waves with radius $R_{\text{wall}%
}=10+\epsilon$ lattice units.}%
\label{free_r10}%
\end{center}
\end{figure}

In Fig. \ref{free_r10} we show the free two-particle energy spectrum in the
center-of-mass frame. \ We have set the spherical wall so that the amplitude
for two-particle separation greater than $10\ $lattice units is strongly
suppressed by a large potential energy due to the spherical wall. \ We can
write this as%
\begin{equation}
V_{0}(\vec{n}_{1}-\vec{n}_{2})\rightarrow V_{0}(\vec{n}_{1}-\vec{n}%
_{2})+V_{\text{wall}}\times\theta(\left\vert \vec{n}_{1}-\vec{n}%
_{2}\right\vert -(10+\epsilon)),
\end{equation}
where $\theta$ is the unit step function and $\epsilon$ is a small positive
number. \ This choice for the spherical wall radius is large enough to probe
wavelengths much larger than the lattice spacing.%
The breaking of rotational invariance due to lattice regularization can be
seen in the small splitting of the different SO$(3,Z)$ representations
comprising each orbital angular momentum multiplet. \ The elements of each
SO$(3,Z)$ representation however remain exactly degenerate.

\begin{figure}
[ptb]
\begin{center}
\includegraphics[
trim=0.000000in 0.000000in 0.000000in 5.019205in,
height=3.0191in,
width=4.2229in
]%
{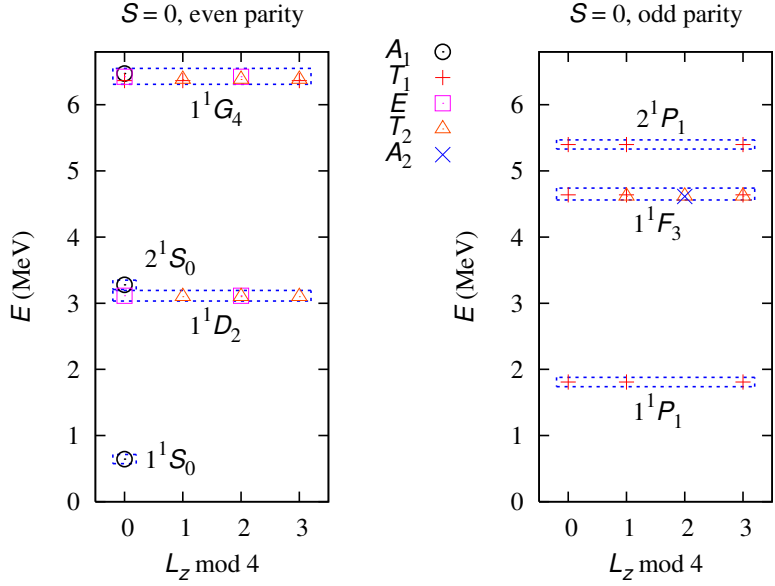}%
\caption{Interacting spectrum for $S=0$ standing waves with $R_{\text{wall}%
}=10+\epsilon$ lattice units.}%
\label{s0_r10}%
\end{center}
\end{figure}

In Fig. \ref{s0_r10} we show the interacting two-particle energy spectrum for
$S=0$ for the same wall radius. \
From the lattice data we find that the interaction is attractive in each of
the $S=0$ channels. \ This is as one might expect since $V_{0}$ is negative
definite. \ To illustrate the calculation of phase shifts, we take for example
the $1^{1}S_{0}$ state. \ For the free particle system the $1^{1}S_{0}$ state
has energy $0.9280$ MeV, which corresponds with a continuum momentum of%
\begin{equation}
k_{\text{free}}=29.52\text{ MeV.}%
\end{equation}
The $1^{1}S_{0}$ state has exactly one extremum and so the free wavefunction
vanishes when%
\begin{equation}
j_{0}(k_{\text{free}}R_{\text{wall}})=0,
\end{equation}%
\begin{equation}
R_{\text{wall}}=\frac{\pi}{k_{\text{free}}}=0.1064\text{ MeV}^{-1}.
\end{equation}
This corresponds with $10.64$ lattice units, and is consistent with our wall
starting at lattice distances greater than $10\ $lattice units. \ For the
interacting system the $1^{1}S_{0}$ state has energy $0.6445$ MeV. \ This
corresponds with a continuum momentum of
\begin{equation}
k=24.60\text{ MeV.}%
\end{equation}
Using (\ref{simple_phaseshift}) we find a $^{1}S_{0}$ phase shift at $k=24.60$
MeV equal to%
\begin{equation}
\delta(^{1}S_{0})=\tan^{-1}\left[  \frac{j_{0}(kR_{\text{wall}})}%
{y_{0}(kR_{\text{wall}})}\right]  =30.0%
{{}^\circ}%
.
\end{equation}
We proceed in this manner for all of the $S=0$ phase shifts. \ In cases where
the spin multiplet is slightly non-degenerate due to the lattice we use the
average energy over all elements of the multiplet.

\begin{figure}
[ptb]
\begin{center}
\includegraphics[
height=6.3624in,
width=4.2108in
]%
{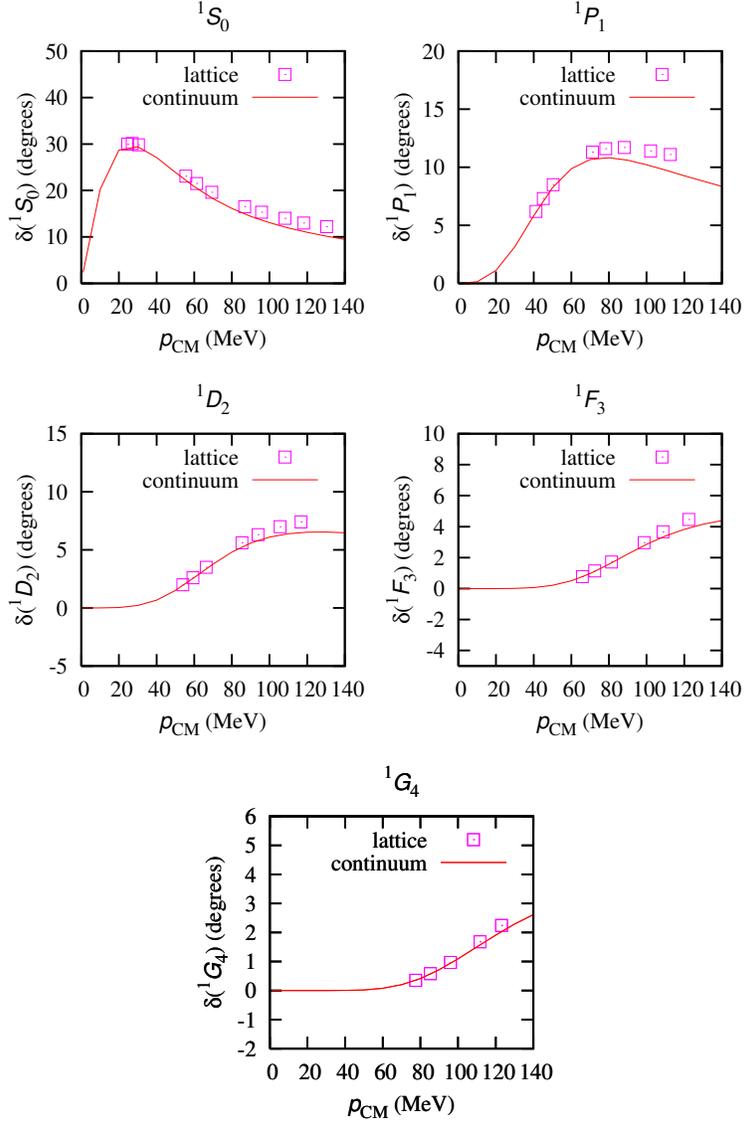}%
\caption{$S=0$ phase shifts for $J\leq4$. \ In order of increasing momentum
the lattice data corresponds with the first radial excitation for
$R_{\text{wall}}=10+\epsilon,9+\epsilon,$ and $8+\epsilon$; second radial
excitation for $R_{\text{wall}}=10+\epsilon,9+\epsilon,$ and $8+\epsilon;$ and
so on.}%
\label{singlet}%
\end{center}
\end{figure}

In Fig. \ref{singlet} we compare lattice results and exact continuum results
for the $^{1}S_{0}$, $^{1}P_{1},$ $^{1}D_{2},$ $^{1}F_{3}$, and $^{1}G_{4}$
phase shifts as a function of the center-of-mass momentum $p_{\text{CM}}%
$.\ \ For the lattice results we use $R_{\text{wall}}=10+\epsilon$,
$9+\epsilon$, and $8+\epsilon$ lattice units. \ In order of increasing
momentum, the lattice data corresponds with the first radial excitation for
$R_{\text{wall}}=10+\epsilon,9+\epsilon,$ and $8+\epsilon$; second radial
excitation of $R_{\text{wall}}=10+\epsilon,9+\epsilon,$ and $8+\epsilon;$ and
so on. \ The continuum results are calculated by solving the
Lippmann-Schwinger equation in momentum space.%

The lattice results are within a few percent of exact results for momenta
below $80$ MeV. \ The error increases to $10\%$ or $15\%$ for momenta near
$120$ MeV. \ This is as good as can be expected without fine-tuning of the
lattice action. \ The lattice spacing $a=(100$ MeV)$^{-1}$ corresponds with a
momentum cutoff equal to $\pi/a=314$ MeV. \ The quality of the results at
higher spin is remarkable considering that $J_{z}$ can only be determined
modulo $4$ on the lattice using (\ref{Jz}).

\section{Lattice calculation for $S=1$}

\begin{figure}
[ptb]
\begin{center}
\includegraphics[
trim=0.000000in 0.000000in 0.000000in 5.019205in,
height=3.0191in,
width=4.2229in
]%
{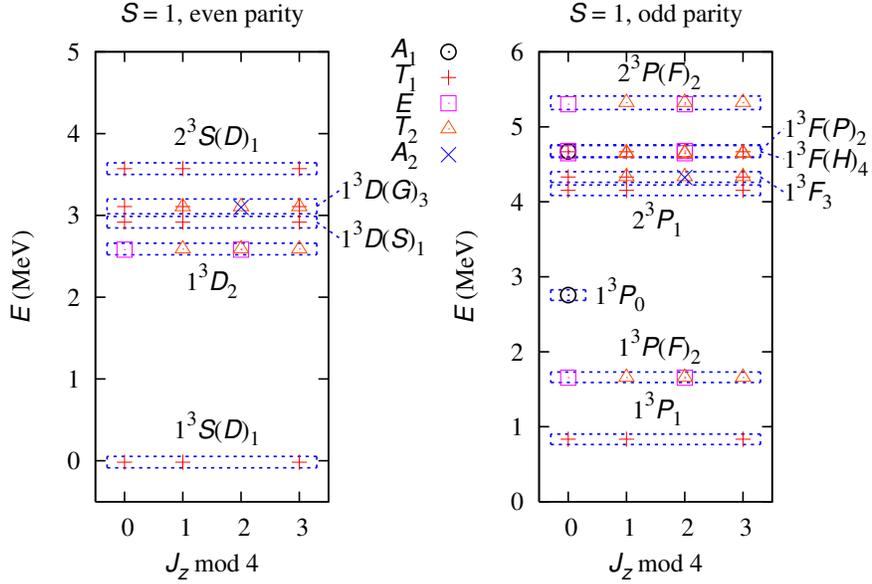}%
\caption{Interacting spectrum for $S=1$ standing waves with $R_{\text{wall}%
}=10+\epsilon$ lattice units.}%
\label{s1_r10}%
\end{center}
\end{figure}

For intrinsic spin $S=1$ the tensor interaction in $V(\vec{r}\,)$ must be
included. \ We define the spin densities,%
\begin{equation}
\rho_{l}^{a^{\dag},a}(\vec{n})=\sum_{i,j=\uparrow,\downarrow}a_{i}^{\dagger
}(\vec{n})\left[  \sigma_{l}\right]  _{ij}a_{j}(\vec{n})\qquad l=1,2,3.
\end{equation}
The transfer matrix has the form%
\begin{align}
M  &  \equiv\colon\exp\left[  -H_{\text{free}}\alpha_{t}-\frac{\alpha_{t}}%
{2}\sum_{\vec{n}_{1},\vec{n}_{2}}V_{0}(\vec{n}_{1}-\vec{n}_{2})\rho^{a^{\dag
},a}(\vec{n}_{1})\rho^{a^{\dag},a}(\vec{n}_{2})\right. \nonumber\\
&  \qquad\qquad\left.  -\frac{\alpha_{t}}{2R_{0}^{2}}\sum_{\vec{n}_{1},\vec
{n}_{2}}\sum_{l,l^{\prime}=1,2,3}V_{0}(\vec{n}_{1}-\vec{n}_{2})T_{ll^{\prime}%
}(\vec{n}_{1}-\vec{n}_{2})\rho_{l}^{a^{\dag},a}(\vec{n}_{1})\rho_{l^{\prime}%
}^{a^{\dag},a}(\vec{n}_{2})\right]  \colon,
\end{align}
where%
\begin{equation}
T_{ll^{\prime}}(\vec{n})=3n_{l}n_{l^{\prime}}-\left\vert \vec{n}\right\vert
^{2}\delta_{ll^{\prime}}.
\end{equation}
In Fig. \ref{s1_r10} we show the interacting two-particle energy spectrum for
$S=1$ for wall radius $R_{\text{wall}}=10+\epsilon$ lattice units.%

The total angular momentum $J$ multiplets are deduced from the approximate
degeneracy of SO$(3,Z)$ representations comprising the multiplet
decompositions in Table \ref{decomp}. \ In some cases the accidental
degeneracy of different $J$ multiplets makes this process difficult. \ For
example the$\ 1^{3}F(P)_{2}$ and $1^{3}F(H)_{4}$ multiplets are nearly
degenerate for $R_{\text{wall}}=10+\epsilon$, as can be seen in Fig.
\ref{s1_r10}. \ In such cases further information can be extracted by
calculating the inner product of the standing wave with spherical harmonics
$Y_{L,L_{z}}(\theta,\phi)$. \ In the infinite volume limit the $1^{3}S(D)_{1}$
bound state has an energy of $-0.170$ MeV, in good agreement with the exact
result $-0.155$ MeV.

\begin{figure}
[ptb]
\begin{center}
\includegraphics[
height=6.3624in,
width=4.2108in
]%
{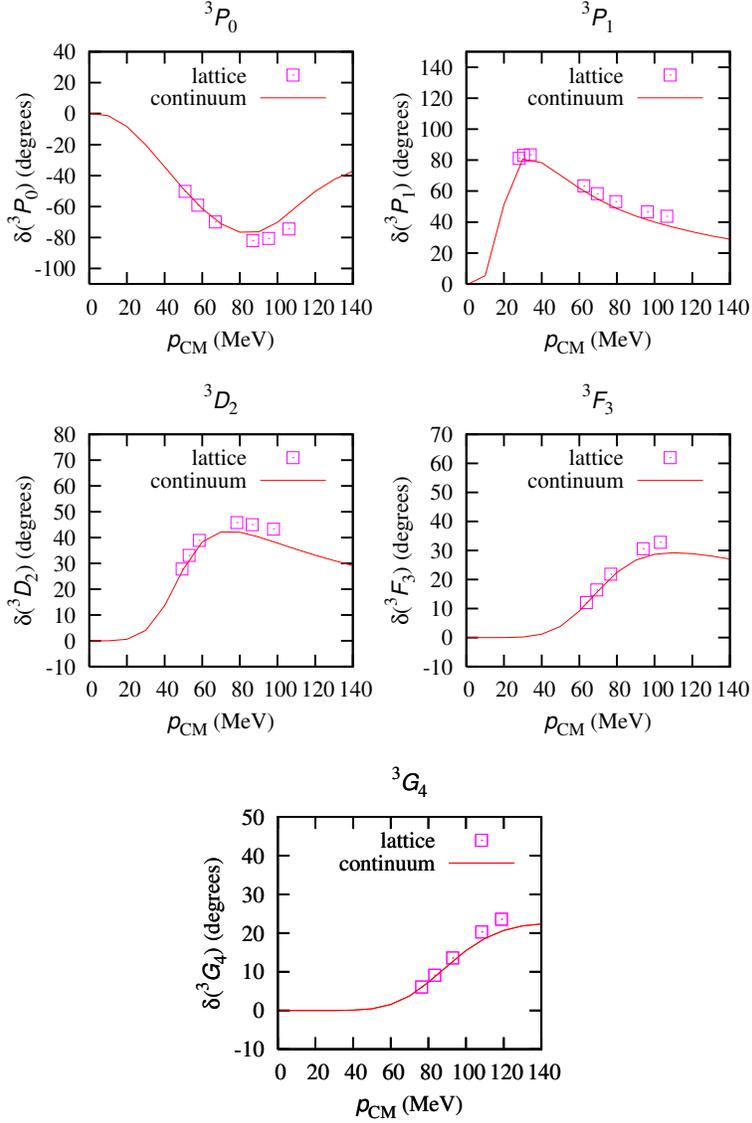}%
\caption{Uncoupled $S=1$ phase shifts for $J\leq4$. \ The lattice data
corresponds with $R_{\text{wall}}=10+\epsilon,9+\epsilon,$ and $8+\epsilon$.}%
\label{triplet_uncoupled}%
\end{center}
\end{figure}

For $S=1$ we start with the uncoupled channels. \ Once the $^{2S+1}L_{J}$
multiplets are identified the process is exactly the same as for the $S=0$
case. \ In Fig. \ref{triplet_uncoupled} we compare lattice results and exact
continuum results for the $^{3}P_{0}$, $^{3}P_{1},$ $^{3}D_{2},$ $^{3}F_{3}$,
and $^{3}G_{4}$ phase shifts. \ For the lattice results we again use
$R_{\text{wall}}=10+\epsilon$, $9+\epsilon$, and $8+\epsilon$ lattice units.%

The tensor interaction produces a strong repulsion in the $^{3}P_{0}$ channel,
enough to overcome the attraction from the central force interaction. \ As in
the $S=0$ case, the lattice results for the uncoupled $S=1$ channels are
within a few percent of exact results for momenta below $80$ MeV and within
$10\%$ or $15\%$ for momenta near $120$ MeV.

\section{Partial wave mixing for $S=1$ at asymptotically large radius}

In the next few sections we consider partial-wave mixing between the partial
waves $L=J-1$ and $L=J+1$ using the Stapp parameterization \cite{Stapp:1956mz}%
. \ It is convenient to use the two-component notation,
\begin{equation}
\left[
\begin{array}
[c]{c}%
R_{J-1}(r)\\
R_{J+1}(r)
\end{array}
\right]  , \label{twocomponent}%
\end{equation}
for the radial part of the wavefunction in continuous space. \ The full
expression is%
\begin{align}
&  R_{J-1}(r)\sum_{L_{z}=-(J-1)}^{J-1}\sum_{S_{z}=-1}^{1}Y_{J-1,L_{z}}%
(\theta,\phi)\left\langle J-1,L_{z};1,S_{z}\right.  \left\vert J,J_{z}%
\right\rangle \times\left\vert 1,S_{z}\right\rangle \nonumber\\
&  +R_{J+1}(r)\sum_{L_{z}=-(J+1)}^{J+1}\sum_{S_{z}=-1}^{1}Y_{J+1,L_{z}}%
(\theta,\phi)\left\langle J+1,L_{z};1,S_{z}\right.  \left\vert J,J_{z}%
\right\rangle \times\left\vert 1,S_{z}\right\rangle ,
\end{align}
where $\left\langle L,L_{z};S,S_{z}\right.  \left\vert J,J_{z}\right\rangle $
denotes the usual Clebsch-Gordon coefficient for adding orbital angular
momenta and intrinsic spin. \ Using this shorthand notation the $\mathbf{S}%
$-matrix can be parametrized as a $2\times2$ matrix of the form%
\begin{equation}
\mathbf{S}=\left[
\begin{array}
[c]{cc}%
e^{i\delta_{J-1}} & 0\\
0 & e^{i\delta_{J+1}}%
\end{array}
\right]  \left[
\begin{array}
[c]{cc}%
\cos2\varepsilon_{J} & i\sin2\varepsilon_{J}\\
i\sin2\varepsilon_{J} & \cos2\varepsilon_{J}%
\end{array}
\right]  \left[
\begin{array}
[c]{cc}%
e^{i\delta_{J-1}} & 0\\
0 & e^{i\delta_{J+1}}%
\end{array}
\right]  .
\end{equation}
Let the incoming wave at asymptotically large radius $r$ be%
\begin{equation}
\Psi_{\text{in}}=-\frac{e^{-i(kr-J\pi/2)}}{2ikr}\left[
\begin{array}
[c]{cc}%
e^{-i\pi/2} & 0\\
0 & e^{i\pi/2}%
\end{array}
\right]  \left[
\begin{array}
[c]{c}%
C\\
D
\end{array}
\right]  .
\end{equation}
Then the outgoing wave at large $r$ is \cite{Blatt:1952, Blatt:1952RMP,
Stapp:1956mz}%
\begin{align}
\Psi_{\text{out}}  &  =\frac{e^{i(kr-J\pi/2)}}{2ikr}\left[
\begin{array}
[c]{cc}%
e^{i\pi/2} & 0\\
0 & e^{-i\pi/2}%
\end{array}
\right]  \mathbf{S}\left[
\begin{array}
[c]{c}%
C\\
D
\end{array}
\right] \nonumber\\
&  =\frac{e^{i(kr-J\pi/2)}}{2ikr}\left[
\begin{array}
[c]{c}%
Ce^{i\pi/2}e^{2i\delta_{J-1}}\cos2\varepsilon_{J}+iDe^{i\pi/2}e^{i\left(
\delta_{J-1}+\delta_{J+1}\right)  }\sin2\varepsilon_{J}\\
De^{-i\pi/2}e^{2i\delta_{J+1}}\cos2\varepsilon_{J}+iCe^{-i\pi/2}e^{i\left(
\delta_{J-1}+\delta_{J+1}\right)  }\sin2\varepsilon_{J}%
\end{array}
\right]  .
\end{align}

It is convenient to define another unitary operator,%
\begin{equation}
\mathbf{W}=\left[
\begin{array}
[c]{cc}%
e^{-i\delta_{J-1}} & 0\\
0 & e^{-i\delta_{J+1}}%
\end{array}
\right]  \left[
\begin{array}
[c]{cc}%
\cos\varepsilon_{J} & -i\sin\varepsilon_{J}\\
-i\sin\varepsilon_{J} & \cos\varepsilon_{J}%
\end{array}
\right]  ,
\end{equation}
so that%
\begin{equation}
\mathbf{W}^{\dagger}=\left[
\begin{array}
[c]{cc}%
\cos\varepsilon_{J} & i\sin\varepsilon_{J}\\
i\sin\varepsilon_{J} & \cos\varepsilon_{J}%
\end{array}
\right]  \left[
\begin{array}
[c]{cc}%
e^{i\delta_{J-1}} & 0\\
0 & e^{i\delta_{J+1}}%
\end{array}
\right]  ,
\end{equation}%
\begin{equation}
\mathbf{W}^{\ast}=\left[
\begin{array}
[c]{cc}%
e^{i\delta_{J-1}} & 0\\
0 & e^{i\delta_{J+1}}%
\end{array}
\right]  \left[
\begin{array}
[c]{cc}%
\cos\varepsilon_{J} & i\sin\varepsilon_{J}\\
i\sin\varepsilon_{J} & \cos\varepsilon_{J}%
\end{array}
\right]  .
\end{equation}
These satisfy the identities%
\begin{equation}
\mathbf{WW}^{\dagger}=\mathbf{W}^{\dagger}\mathbf{W}=1,
\end{equation}%
\begin{equation}
\mathbf{W}^{\ast}\mathbf{W}^{\dagger}=\mathbf{S}.
\end{equation}
To construct a real-valued standing wave, we let the asymptotic incoming wave
be%
\begin{equation}
-\frac{e^{-i(kr-J\pi/2)}}{2ikr}\left[
\begin{array}
[c]{cc}%
e^{-i\pi/2} & 0\\
0 & e^{i\pi/2}%
\end{array}
\right]  \mathbf{W}\left[
\begin{array}
[c]{c}%
C\\
D
\end{array}
\right]
\end{equation}
for real numbers $C$ and $D$. \ Then the outgoing wave is%
\begin{equation}
\frac{e^{i(kr-J\pi/2)}}{2ikr}\left[
\begin{array}
[c]{cc}%
e^{i\pi/2} & 0\\
0 & e^{-i\pi/2}%
\end{array}
\right]  \mathbf{SW}\left[
\begin{array}
[c]{c}%
C\\
D
\end{array}
\right]  =\frac{e^{i(kr-J\pi/2)}}{2ikr}\left[
\begin{array}
[c]{cc}%
e^{i\pi/2} & 0\\
0 & e^{-i\pi/2}%
\end{array}
\right]  \mathbf{W}^{\ast}\left[
\begin{array}
[c]{c}%
C\\
D
\end{array}
\right]  .
\end{equation}
The resulting standing wave is then%
\begin{align}
\Psi &  =2\operatorname{Re}\left\{  \frac{e^{i(kr-J\pi/2)}}{2ikr}\left[
\begin{array}
[c]{cc}%
e^{i\pi/2} & 0\\
0 & e^{-i\pi/2}%
\end{array}
\right]  \mathbf{W}^{\ast}\left[
\begin{array}
[c]{c}%
C\\
D
\end{array}
\right]  \right\} \nonumber\\
&  =\frac{1}{kr}\left[
\begin{array}
[c]{c}%
C\sin\left(  kr-\frac{J-1}{2}\pi+\delta_{J-1}\right)  \cos\varepsilon
_{J}+D\cos\left(  kr-\frac{J-1}{2}\pi+\delta_{J-1}\right)  \sin\varepsilon
_{J}\\
D\sin\left(  kr-\frac{J+1}{2}\pi+\delta_{J+1}\right)  \cos\varepsilon
_{J}+C\cos\left(  kr-\frac{J+1}{2}\pi+\delta_{J+1}\right)  \sin\varepsilon_{J}%
\end{array}
\right]  .
\end{align}

We now impose a hard spherical wall boundary at $r=R_{\text{wall}}$. \ Both
partial waves vanish at the wall, and we define the angle $\Delta$ so that%
\begin{equation}
-\Delta=kR_{\text{wall}}-\frac{J-1}{2}\pi.
\end{equation}
Then%
\begin{equation}
\frac{C}{D}\tan\left(  -\Delta+\delta_{J-1}\right)  =-\tan\varepsilon_{J}%
\end{equation}
and%
\begin{equation}
\frac{D}{C}\tan\left(  -\Delta-\pi+\delta_{J+1}\right)  =\frac{D}{C}%
\tan\left(  -\Delta+\delta_{J+1}\right)  =-\tan\varepsilon_{J}.
\end{equation}
In general there are two solutions for $\Delta$ per angular interval $\pi$.
\ In the asymptotic case where $R_{\text{wall}}\gg k^{-1}$ the spacing between
energy levels becomes infinitesimal. \ We can therefore choose two independent
solutions $\Delta^{I}$ and $\Delta^{II}$ with nearly the same values of $k$,
and we neglect the difference in $k$ in the following steps.

The solutions $\Delta^{I}$ and $\Delta^{II}$ satisfy%
\begin{equation}
\tan\left(  -\Delta^{I,II}+\delta_{J-1}\right)  \tan\left(  -\Delta
^{I,II}+\delta_{J+1}\right)  =\tan^{2}\varepsilon_{J}. \label{constraint1}%
\end{equation}
The standing wave now looks like%
\begin{align}
\Psi &  =\frac{1}{kr}\left[
\begin{array}
[c]{c}%
C\sin\left(  kr-\frac{J-1}{2}\pi+\delta_{J-1}\right)  \cos\varepsilon
_{J}-C\frac{\sin\left(  -\Delta+\delta_{J-1}\right)  }{\cos\left(
-\Delta+\delta_{J-1}\right)  }\cos\left(  kr-\frac{J-1}{2}\pi+\delta
_{J-1}\right)  \cos\varepsilon_{J}\\
-C\sin\left(  kr-\frac{J+1}{2}\pi+\delta_{J+1}\right)  \frac{\cos\left(
-\Delta+\delta_{J+1}\right)  }{\sin\left(  -\Delta+\delta_{J+1}\right)  }%
\sin\varepsilon_{J}+C\cos\left(  kr-\frac{J+1}{2}\pi+\delta_{J+1}\right)
\sin\varepsilon_{J}%
\end{array}
\right] \nonumber\\
&  =\frac{C}{kr}\left[
\begin{array}
[c]{c}%
\frac{\cos\varepsilon_{J}}{\cos\left(  -\Delta+\delta_{J-1}\right)  }%
\sin\left[  kr-\frac{J-1}{2}\pi+\Delta\right] \\
-\frac{\sin\varepsilon_{J}}{\sin\left(  -\Delta+\delta_{J+1}\right)  }%
\sin\left[  kr-\frac{J+1}{2}\pi+\Delta\right]
\end{array}
\right]  .
\end{align}
There are two linearly independent real-valued standing wave solutions.
\ We\ write them as%
\begin{equation}
\Psi^{I,II}\propto\frac{1}{kr}\left[
\begin{array}
[c]{c}%
A_{J-1}^{I,II}\sin\left(  kr-\frac{J-1}{2}\pi+\Delta^{I,II}\right) \\
A_{J+1}^{I,II}\sin\left(  kr-\frac{J+1}{2}\pi+\Delta^{I,II}\right)
\end{array}
\right]  .
\end{equation}
Then%
\begin{equation}
A_{J-1}^{I,II}\tan\varepsilon_{J}=-A_{J+1}^{I,II}\frac{\sin\left(
-\Delta^{I,II}+\delta_{J+1}\right)  }{\cos\left(  -\Delta^{I,II}+\delta
_{J-1}\right)  }. \label{constraint2}%
\end{equation}

Only three of the constraints in (\ref{constraint1}) and (\ref{constraint2})
are needed to determine the parameters $\delta_{J-1}$, $\delta_{J+1},$ and
$\varepsilon_{J}$. \ Therefore one constraint, for example%
\begin{equation}
A_{J-1}^{II}\tan\varepsilon_{J}=-A_{J+1}^{II}\frac{\sin\left(  -\Delta
^{II}+\delta_{J+1}\right)  }{\cos\left(  -\Delta^{II}+\delta_{J-1}\right)  },
\label{constraint4}%
\end{equation}
must be redundant. \ This makes sense in light of the symmetry associated with
the interchange of the two solutions $\Delta^{I}$ and $\Delta^{II}$. \ The
orthogonality of $\Psi^{I}$ and $\Psi^{II}$ implies%
\begin{equation}
A_{J-1}^{I}A_{J-1}^{II}+A_{J+1}^{I}A_{J+1}^{II}=0,
\end{equation}%
\begin{equation}
\frac{A_{J+1}^{I}}{A_{J-1}^{I}}=-\frac{A_{J-1}^{II}}{A_{J+1}^{II}}.
\label{orthogonality}%
\end{equation}
From this one can derive (\ref{constraint4}) from (\ref{constraint1}) and
(\ref{constraint2}).

\section{Expansion for small $\varepsilon_{J}$}

For our test potential we have chosen the tensor force to be very strong, and
so the mixing angles are quite large. \ In several real physical systems
however, for example nucleon-nucleon scattering at low momenta, the mixing
angles are quite small. \ For small $\varepsilon_{J}$ it is useful to expand
around $\varepsilon_{J}=0$. \ We choose $\Delta^{I}$ to be the solution that
equals $\delta_{J-1}$ at zero mixing$,$ and $\Delta^{II}$ to be the solution
that equals $\delta_{J+1}$ at zero mixing. \ Since
\begin{equation}
\tan\left(  -\Delta^{I}+\delta_{J-1}\right)  \tan\left(  -\Delta^{I}%
+\delta_{J+1}\right)  =\tan^{2}\varepsilon_{J},
\end{equation}
the first correction to $\Delta^{I}$ and $\Delta^{II}$ comes at order
$O(\varepsilon_{J}^{2})$,%
\begin{align}
\Delta^{I}  &  =\delta_{J-1}+c^{I}\varepsilon_{J}^{2}+O(\varepsilon_{J}%
^{4}),\\
\Delta^{II}  &  =\delta_{J+1}+c^{II}\varepsilon_{J}^{2}+O(\varepsilon_{J}%
^{4}).
\end{align}
We find%
\begin{equation}
-c^{I}\varepsilon_{J}^{2}\tan\left(  \delta_{J+1}-\delta_{J-1}\right)
=\varepsilon_{J}^{2}+O(\varepsilon_{J}^{4}),
\end{equation}%
\begin{equation}
c^{I}=-\frac{1}{\tan\left(  \delta_{J+1}-\delta_{J-1}\right)  }.
\end{equation}
Similarly%
\begin{equation}
-c^{II}\varepsilon_{J}^{2}\tan\left(  \delta_{J-1}-\delta_{J+1}\right)
=\varepsilon_{J}^{2}+O(\varepsilon_{J}^{4}),
\end{equation}%
\begin{equation}
c^{II}=\frac{1}{\tan\left(  \delta_{J+1}-\delta_{J-1}\right)  }.
\end{equation}

Since%
\begin{equation}
A_{J-1}^{I}\tan\varepsilon_{J}=-A_{J+1}^{I}\frac{\sin\left(  -\Delta
^{I}+\delta_{J+1}\right)  }{\cos\left(  -\Delta^{I}+\delta_{J-1}\right)  },
\end{equation}
we have%
\begin{equation}
\sin\left(  \delta_{J+1}-\delta_{J-1}\right)  =-\frac{A_{J-1}^{I}%
\varepsilon_{J}}{A_{J+1}^{I}}+O(\varepsilon_{J}^{2}).
\end{equation}
Up to $O(\varepsilon_{J}^{2})$ then%
\begin{equation}
\sin\left(  \Delta^{II}-\Delta^{I}\right)  =-\frac{A_{J-1}^{I}\varepsilon_{J}%
}{A_{J+1}^{I}}+O(\varepsilon_{J}^{2}),
\end{equation}
or%
\begin{equation}
\varepsilon_{J}=-\frac{A_{J+1}^{I}}{A_{J-1}^{I}}\sin\left(  \Delta^{II}%
-\Delta^{I}\right)  +O(\varepsilon_{J}^{3}). \label{smalleps}%
\end{equation}

Similarly we have%
\begin{equation}
A_{J-1}^{II}\tan\varepsilon_{J}=-A_{J+1}^{II}\frac{\sin\left(  -\Delta
^{II}+\delta_{J+1}\right)  }{\cos\left(  -\Delta^{II}+\delta_{J-1}\right)  },
\end{equation}
and so%
\begin{align}
A_{J-1}^{II}  &  =-A_{J+1}^{II}\frac{-c^{II}\varepsilon_{J}^{2}}%
{\varepsilon_{J}\cos\left(  -\delta_{J+1}+\delta_{J-1}\right)  }%
+O(\varepsilon_{J}^{3})\nonumber\\
&  =A_{J+1}^{II}\frac{\varepsilon_{J}}{\sin\left(  \delta_{J+1}-\delta
_{J-1}\right)  }+O(\varepsilon_{J}^{3}),
\end{align}
or%
\begin{equation}
\sin\left(  \delta_{J+1}-\delta_{J-1}\right)  =\frac{A_{J+1}^{II}%
\varepsilon_{J}}{A_{J-1}^{II}}+O(\varepsilon_{J}^{2}).
\end{equation}
Up to order $O(\varepsilon_{J}^{2})$ we find%
\begin{equation}
\varepsilon_{J}=\frac{A_{J-1}^{II}}{A_{J+1}^{II}}\sin\left(  \Delta
^{II}-\Delta^{I}\right)  +O(\varepsilon_{J}^{3}).
\end{equation}
This is consistent with (\ref{smalleps}) given the orthogonality condition,%
\begin{equation}
\frac{A_{J+1}^{I}}{A_{J-1}^{I}}=-\frac{A_{J-1}^{II}}{A_{J+1}^{II}}\,.
\end{equation}

\section{Hard spherical wall at non-asymptotic radius}

The constraints in (\ref{constraint1}) and (\ref{constraint2}) hold if the
spherical wall is at asymptotically large radius, $kR_{\text{wall}}\gg1$. \ In
this limit the nodes of the spherical Bessel functions $j_{J-1}(kr)$ and
$j_{J+1}(kr)$ coincide, and we determine one shift angle for both partial
waves,%
\begin{equation}
-\Delta=kR_{\text{wall}}-\frac{J-1}{2}\pi.
\end{equation}
However for numerical calculations it is more convenient to work with smaller
values of $kR_{\text{wall}}$. \ In this case the coincidence of the nodes for
$L=J-1$ and $L=J+1$ at the wall boundary does not automatically imply the
coincidence of nodes for $kr\gg1$. \ Nevertheless we can still work with the
asymptotic form for $kr\gg1$ if the node remains at $R_{\text{wall}}$ but the
spherical wall is removed,%
\begin{equation}
\Psi=\frac{1}{kr}\left[
\begin{array}
[c]{c}%
C\sin\left(  kr-\frac{J-1}{2}\pi+\delta_{J-1}\right)  \cos\varepsilon
_{J}+D\cos\left(  kr-\frac{J-1}{2}\pi+\delta_{J-1}\right)  \sin\varepsilon
_{J}\\
D\sin\left(  kr-\frac{J+1}{2}\pi+\delta_{J+1}\right)  \cos\varepsilon
_{J}+C\cos\left(  kr-\frac{J+1}{2}\pi+\delta_{J+1}\right)  \sin\varepsilon_{J}%
\end{array}
\right]  .
\end{equation}
Instead of one angle $\Delta$, we define two angles $\Delta_{J-1}$ and
$\Delta_{J+1}$ so that the $J-1$ partial wave vanishes when%
\begin{equation}
-\Delta_{J-1}=kr-\frac{J-1}{2}\pi,
\end{equation}
and the $J+1$ partial wave vanishes when%
\begin{equation}
-\Delta_{J+1}=kr-\frac{J+1}{2}\pi.
\end{equation}
Then%
\begin{equation}
\frac{C}{D}\tan\left(  -\Delta_{J-1}+\delta_{J-1}\right)  =-\tan
\varepsilon_{J}%
\end{equation}
and%
\begin{equation}
\frac{D}{C}\tan\left(  -\Delta_{J+1}+\delta_{J+1}\right)  =-\tan
\varepsilon_{J}.
\end{equation}
In general for a given $k$, there are two pairs of solutions $\left(
\Delta_{J-1},\Delta_{J+1}\right)  $ per angular interval $\pi$ corresponding
with different positions of the nodes. \ However instead of taking $k$ as the
independent variable, we consider the location of the spherical wall radius as
the independent variable. \ We denote the two solutions as $\left(
\Delta_{J-1}^{I},\Delta_{J+1}^{I},k^{I}\right)  $ and $\left(  \Delta
_{J-1}^{II},\Delta_{J+1}^{II},k^{II}\right)  $. \ We have%
\begin{equation}
\tan\left(  -\Delta_{J-1}^{I}+\delta_{J-1}\right)  \tan\left(  -\Delta
_{J+1}^{I}+\delta_{J+1}\right)  =\tan^{2}\varepsilon_{J},
\end{equation}
at momentum $k^{I}$, and
\begin{equation}
\tan\left(  -\Delta_{J-1}^{II}+\delta_{J-1}\right)  \tan\left(  -\Delta
_{J+1}^{II}+\delta_{J+1}\right)  =\tan^{2}\varepsilon_{J},
\end{equation}
at momentum $k^{II}$.

The standing wave now looks like%
\begin{align}
\Psi &  =\frac{1}{kr}\left[
\begin{array}
[c]{c}%
C\sin\left(  kr-\frac{J-1}{2}\pi+\delta_{J-1}\right)  \cos\varepsilon
_{J}+D\cos\left(  kr-\frac{J-1}{2}\pi+\delta_{J-1}\right)  \sin\varepsilon
_{J}\\
D\sin\left(  kr-\frac{J+1}{2}\pi+\delta_{J+1}\right)  \cos\varepsilon
_{J}+C\cos\left(  kr-\frac{J+1}{2}\pi+\delta_{J+1}\right)  \sin\varepsilon_{J}%
\end{array}
\right] \nonumber\\
&  =\frac{C}{kr}\left[
\begin{array}
[c]{c}%
\frac{\cos\varepsilon_{J}}{\cos\left(  -\Delta_{J-1}+\delta_{J-1}\right)
}\sin\left[  kr-\frac{J-1}{2}\pi+\Delta_{J-1}\right] \\
-\frac{\sin\varepsilon_{J}}{\sin\left(  -\Delta_{J+1}+\delta_{J+1}\right)
}\sin\left[  kr-\frac{J+1}{2}\pi+\Delta_{J+1}\right]
\end{array}
\right]  .
\end{align}
We\ write these as%
\begin{equation}
\Psi^{I}\propto\frac{1}{k^{I}r}\left[
\begin{array}
[c]{c}%
A_{J-1}^{I}\sin\left(  k^{I}r-\frac{J-1}{2}\pi+\Delta_{J-1}^{I}\right) \\
A_{J+1}^{I}\sin\left(  k^{I}r-\frac{J+1}{2}\pi+\Delta_{J+1}^{I}\right)
\end{array}
\right]
\end{equation}
and%
\begin{equation}
\Psi^{II}\propto\frac{1}{k^{II}r}\left[
\begin{array}
[c]{c}%
A_{J-1}^{II}\sin\left(  k^{II}r-\frac{J-1}{2}\pi+\Delta_{J-1}^{II}\right) \\
A_{J+1}^{II}\sin\left(  k^{II}r-\frac{J+1}{2}\pi+\Delta_{J+1}^{II}\right)
\end{array}
\right]  .
\end{equation}
So the generalization of equations (\ref{constraint1}) and (\ref{constraint2})
to the non-asymptotic case is%
\begin{equation}
\tan\left(  -\Delta_{J-1}^{I}+\delta_{J-1}\right)  \tan\left(  -\Delta
_{J+1}^{I}+\delta_{J+1}\right)  =\tan^{2}\varepsilon_{J},
\label{newconstraint1}%
\end{equation}%
\begin{equation}
\tan\left(  -\Delta_{J-1}^{II}+\delta_{J-1}\right)  \tan\left(  -\Delta
_{J+1}^{II}+\delta_{J+1}\right)  =\tan^{2}\varepsilon_{J},
\label{newconstraint2}%
\end{equation}%
\begin{equation}
A_{J-1}^{I}\tan\varepsilon_{J}=-A_{J+1}^{I}\frac{\sin\left(  -\Delta_{J+1}%
^{I}+\delta_{J+1}\right)  }{\cos\left(  -\Delta_{J-1}^{I}+\delta_{J-1}\right)
}, \label{newconstraint3}%
\end{equation}%
\begin{equation}
A_{J-1}^{II}\tan\varepsilon_{J}=-A_{J+1}^{II}\frac{\sin\left(  -\Delta
_{J+1}^{II}+\delta_{J+1}\right)  }{\cos\left(  -\Delta_{J-1}^{II}+\delta
_{J-1}\right)  }. \label{newconstraint4}%
\end{equation}
We note that the phase shifts and mixing angle in (\ref{newconstraint1}) and
(\ref{newconstraint3}) are at momentum $k^{I}$ while the phase shifts and
mixing angle in (\ref{newconstraint2}) and (\ref{newconstraint4}) are at
momentum $k^{II}$. \ In the calculation one must in general interpolate
between values for $k=k^{I}$ and $k=k^{II}$. \ However the need for
interpolation is largely eliminated by considering only close pairs of values
$k^{I}\approx k^{II}$ in solving (\ref{newconstraint1})-(\ref{newconstraint4}%
). \ This can be done for example by considering the $(n+1)^{\text{st}}%
$-radial excitation of $L=J-1$ together with the $n^{\text{th}}$-radial
excitation of $L=J+1$. \ In this scheme we use%
\begin{equation}
\tan\left(  -\Delta_{J-1}^{I}+\delta_{J-1}(k^{I})\right)  \tan\left(
-\Delta_{J+1}^{I}+\delta_{J+1}(k^{I})\right)  =\tan^{2}\left[  \varepsilon
_{J}(k^{I})\right]  , \label{kI_1}%
\end{equation}%
\begin{equation}
\tan\left(  -\Delta_{J-1}^{II}+\delta_{J-1}(k^{I})\right)  \tan\left(
-\Delta_{J+1}^{II}+\delta_{J+1}(k^{I})\right)  \approx\tan^{2}\left[
\varepsilon_{J}(k^{I})\right]  , \label{kI_2}%
\end{equation}%
\begin{equation}
A_{J-1}^{I}\tan\left[  \varepsilon_{J}(k^{I})\right]  =-A_{J+1}^{I}\frac
{\sin\left(  -\Delta_{J+1}^{I}+\delta_{J+1}(k^{I})\right)  }{\cos\left(
-\Delta_{J-1}^{I}+\delta_{J-1}(k^{I})\right)  }, \label{kI_3}%
\end{equation}
for the phase shifts and mixing angle at $k=k^{I}$, and%
\begin{equation}
\tan\left(  -\Delta_{J-1}^{I}+\delta_{J-1}(k^{II})\right)  \tan\left(
-\Delta_{J+1}^{I}+\delta_{J+1}(k^{II})\right)  \approx\tan^{2}\left[
\varepsilon_{J}(k^{II})\right]  , \label{kII_1}%
\end{equation}%
\begin{equation}
\tan\left(  -\Delta_{J-1}^{II}+\delta_{J-1}(k^{II})\right)  \tan\left(
-\Delta_{J+1}^{II}+\delta_{J+1}(k^{II})\right)  =\tan^{2}\left[
\varepsilon_{J}(k^{II})\right]  , \label{kII_2}%
\end{equation}%
\begin{equation}
A_{J-1}^{II}\tan\left[  \varepsilon_{J}(k^{II})\right]  =-A_{J+1}^{II}%
\frac{\sin\left(  -\Delta_{J+1}^{II}+\delta_{J+1}(k^{II})\right)  }%
{\cos\left(  -\Delta_{J-1}^{II}+\delta_{J-1}(k^{II})\right)  }, \label{kII_3}%
\end{equation}
for the phase shifts and mixing angle at $k=k^{II}$. \ This is the technique
we use for the lattice results presented here. \ If more accuracy is required
then this pair technique for $k^{I}\approx k^{II}$ can be used as a starting
point to determine numerical derivatives for the phase shifts and mixing
angle. \ Then the equations (\ref{newconstraint1})-(\ref{newconstraint4}) can
be solved again while keeping all terms at order $O(k^{I}-k^{II})$.

In the limit of small mixing the expressions in (\ref{kI_1})-(\ref{kII_3})
reduce to%
\begin{equation}
\delta_{J-1}(k^{I})=\Delta_{J-1}^{I}+\frac{\varepsilon_{J}^{2}(k^{I})}%
{\tan\left(  -\Delta_{J+1}^{I}+\delta_{J+1}(k^{I})\right)  }+O(\varepsilon
_{J}^{4}),
\end{equation}%
\begin{equation}
\varepsilon_{J}(k^{I})=-\frac{A_{J+1}^{I}}{A_{J-1}^{I}}\sin\left(
\Delta_{J+1}^{II}-\Delta_{J+1}^{I}\right)  +O(\varepsilon_{J}^{3}),
\end{equation}
at $k=k^{I}$ and%
\begin{equation}
\delta_{J+1}(k^{II})=\Delta_{J+1}^{II}+\frac{\varepsilon_{J}^{2}(k^{II})}%
{\tan\left(  -\Delta_{J-1}^{II}+\delta_{J-1}(k^{II})\right)  }+O(\varepsilon
_{J}^{4}),
\end{equation}%
\begin{equation}
\varepsilon_{J}(k^{II})=\frac{A_{J-1}^{II}}{A_{J+1}^{II}}\sin\left(
\Delta_{J-1}^{II}-\Delta_{J-1}^{I}\right)  +O(\varepsilon_{J}^{3}).
\end{equation}
at $k=k^{II}$.

\section{Results for $S=1$ coupled channels}

\begin{figure}
[ptb]
\begin{center}
\includegraphics[
height=4.4987in,
width=4.2047in
]%
{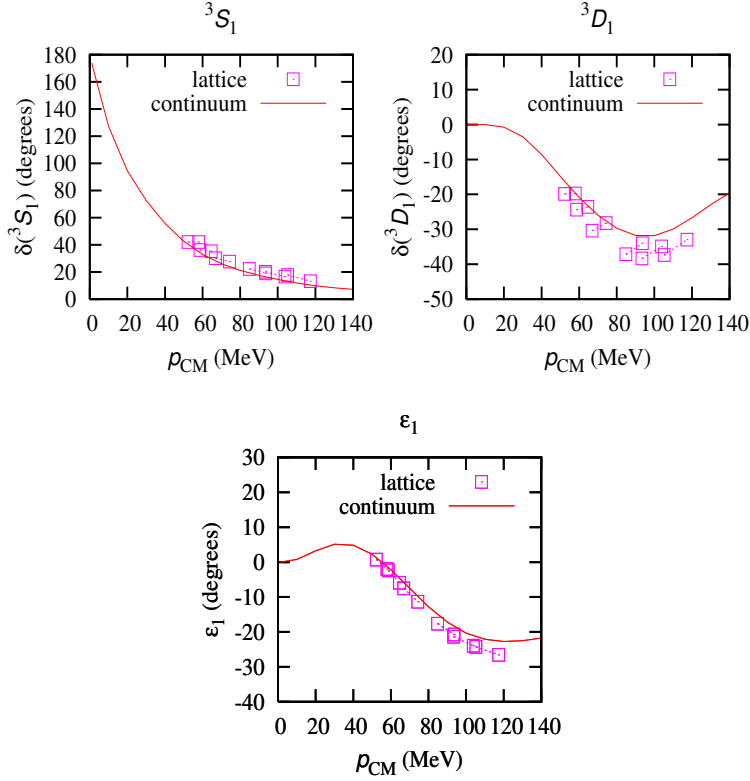}%
\caption{Coupled partial waves and mixing angle for $J=1$. \ The pairs of
points connected by dotted lines indicate pairs of solutions at $k=k^{I}$ and
$k=k^{II}$.}%
\label{j1}%
\end{center}
\end{figure}

\begin{figure}
[ptbptb]
\begin{center}
\includegraphics[
height=4.4987in,
width=4.2047in
]%
{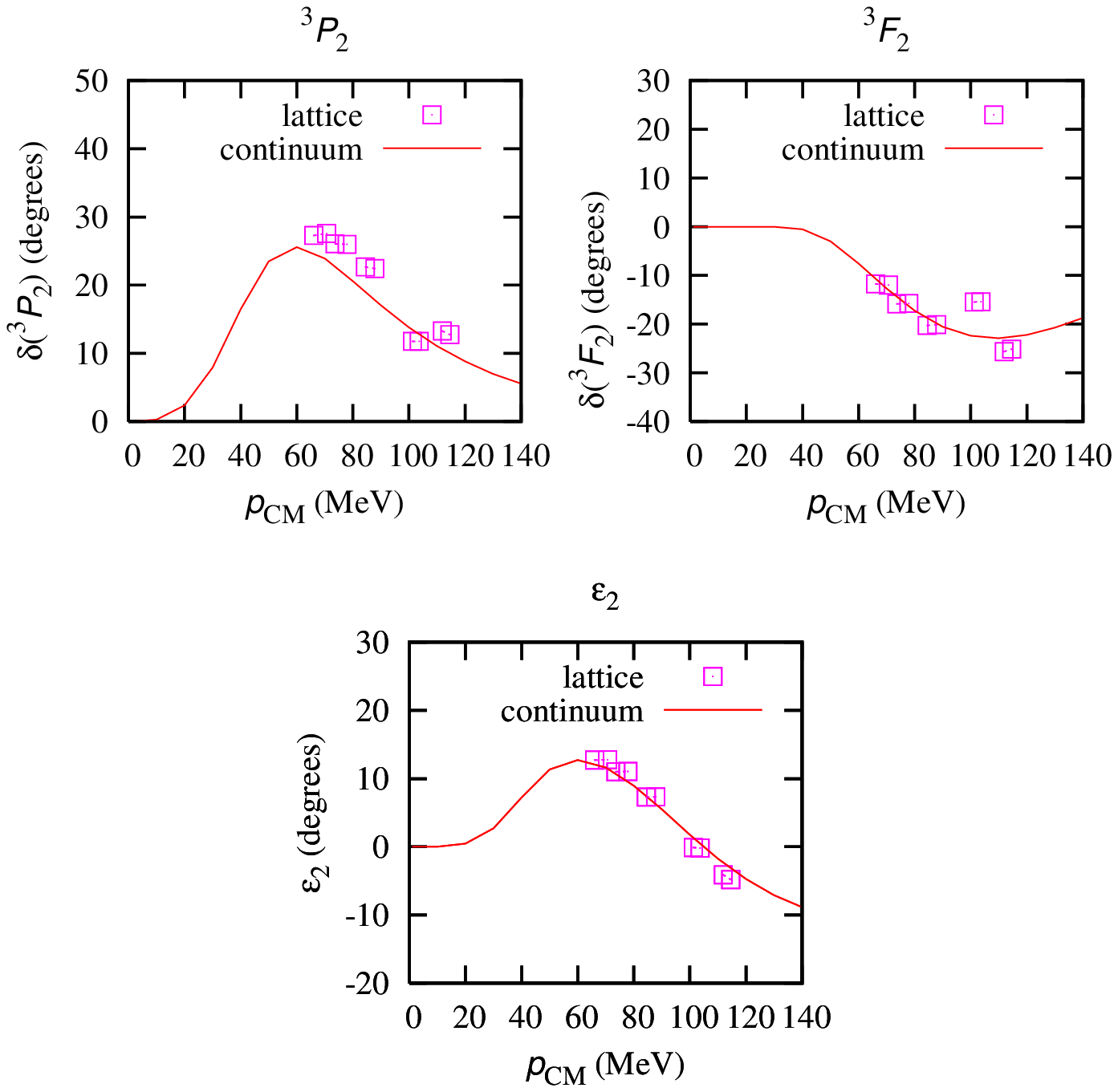}%
\caption{Coupled partial waves and mixing angle for $J=2.$}%
\label{j2}%
\end{center}
\end{figure}

\begin{figure}
[ptbptbptb]
\begin{center}
\includegraphics[
height=4.4987in,
width=4.2047in
]%
{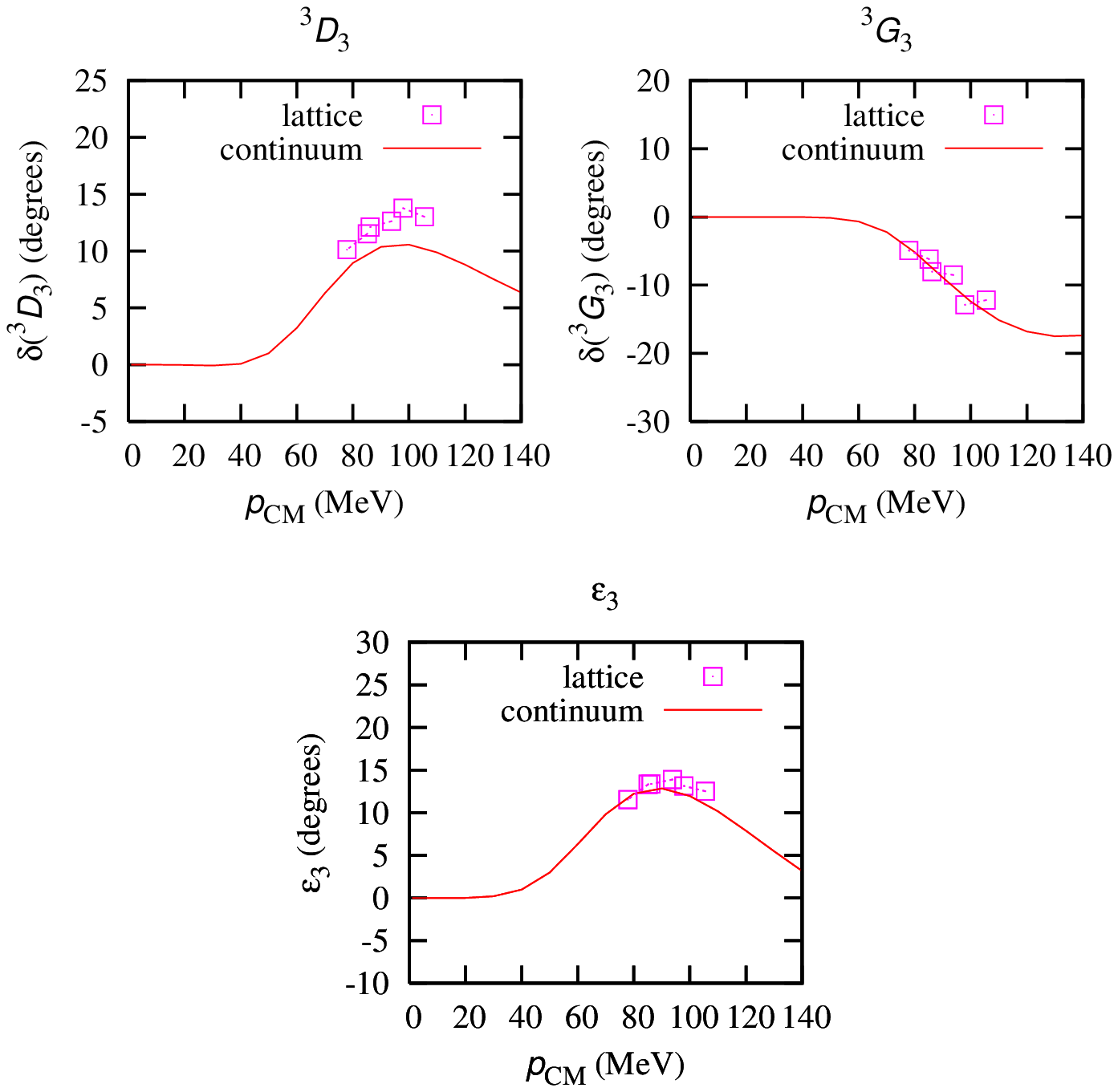}%
\caption{Coupled partial waves and mixing angle for $J=3.$}%
\label{j3}%
\end{center}
\end{figure}

\begin{figure}
[ptbptbptbptb]
\begin{center}
\includegraphics[
height=4.4987in,
width=4.2047in
]%
{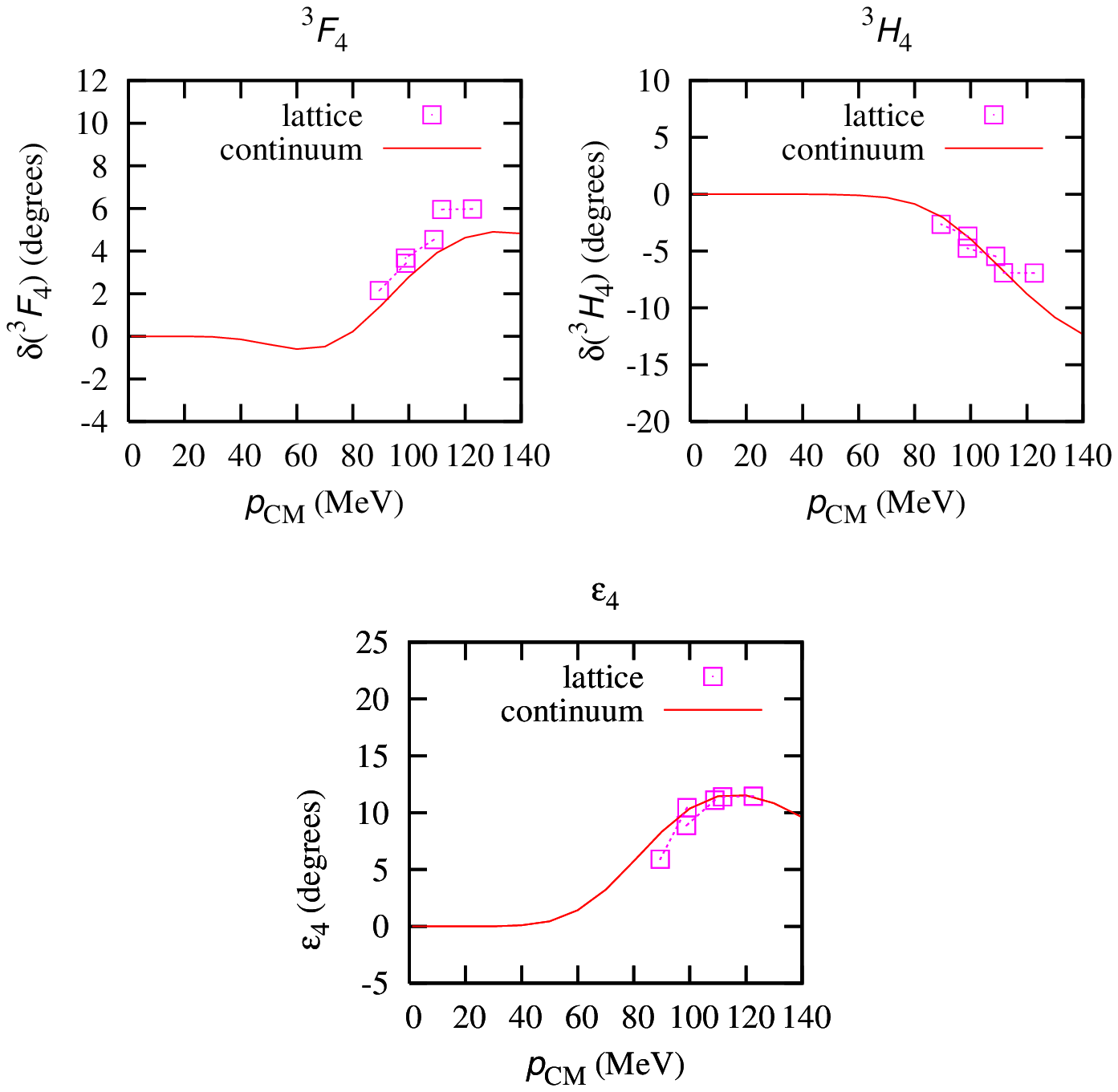}%
\caption{Coupled partial waves and mixing angle for $J=4.$}%
\label{j4}%
\end{center}
\end{figure}

In Fig. \ref{j1} we show lattice and continuum results for $^{3}S_{1}$,
$^{3}D_{1}$ partial waves and $J=1$ mixing angle $\varepsilon_{1}$. \ The
$^{3}P_{2}$, $^{3}F_{2}$ partial waves and mixing angle $\varepsilon_{2}$ are
shown in Fig. \ref{j2}. \ The $^{3}D_{3}$, $^{3}G_{3}$ partial waves and
mixing angle $\varepsilon_{3}$ are shown in Fig. \ref{j3}. \ The $^{3}F_{4}$,
$^{3}H_{4}$ partial waves and mixing angle $\varepsilon_{4}$ are shown in Fig.
\ref{j4}. \ As before we use $R_{\text{wall}}=10+\epsilon$, $9+\epsilon$, and
$8+\epsilon$ lattice units. \ The pairs of points connected by dotted lines
indicate pairs of solutions at $k=k^{I}$ and $k=k^{II}$. \ The partial-wave
ratios%
\begin{equation}
\frac{A_{J-1}^{I}}{A_{J+1}^{I}},\text{ }\frac{A_{J-1}^{II}}{A_{J+1}^{II}},
\end{equation}
are determined by computing the inner product of the standing wave near the
spherical wall with spherical harmonics.%

The lattice results in the coupled channels are somewhat less accurate than
the $S=0$ results and uncoupled $S=1$ results. \ However they are still within
$20\%$ of the exact results up to momentum $120$ MeV. \ The leading error
appears to come from mixing with extraneous channels due to broken rotational
invariance on the lattice. \ For example the $^{3}S_{1}$-$^{3}D_{1}$ standing
waves have a small but non-negligible amount of a $^{3}D_{3}$ component.
\ Nevertheless overall the lattice results are remarkably good for
partial-wave mixing at higher total angular momentum.

\section{Summary and discussion}

We have discussed a general technique for measuring phase shifts and mixing
angles for two-particle scattering on the lattice. \ In the center-of-mass
frame we impose a hard spherical wall at large fixed radius $R_{\text{wall}}$.
\ For channels without mixing we identify total angular momentum $J$
multiplets by the approximate degeneracy of SO$(3,Z)$ representations and
calculate phase shifts from the energies of the spherical standing waves.
\ For channels with partial-wave mixing, further information is extracted by
decomposing the standing wave at the wall boundary into spherical harmonics.
\ The coupled-channels equations are then solved to extract the phase shifts
and mixing angles. \ The method was tested by computing phase shifts and
mixing angles for $J$ less than or equal to $4$ for an attractive Gaussian
potential with both central and tensor force parts. \ The Gaussian envelope
had a characteristic size of $R_{0}=2\times10^{-2}$\ MeV$^{-1}$, and the
strength was tuned to produce a shallow $^{3}S(D)_{1}$ bound state. \ At
spatial lattice spacing $a=(100$ MeV$)^{-1}$ and temporal lattice spacing
$a_{t}=(70$ MeV$)^{-1}$ we found agreement with the exact results at the
$10\%-20\%$ level for momenta up to $120$ MeV.

\begin{figure}
[ptb]
\begin{center}
\includegraphics[
trim=0.000000in 0.000000in 0.000000in 5.018204in,
height=3.0191in,
width=4.2229in
]%
{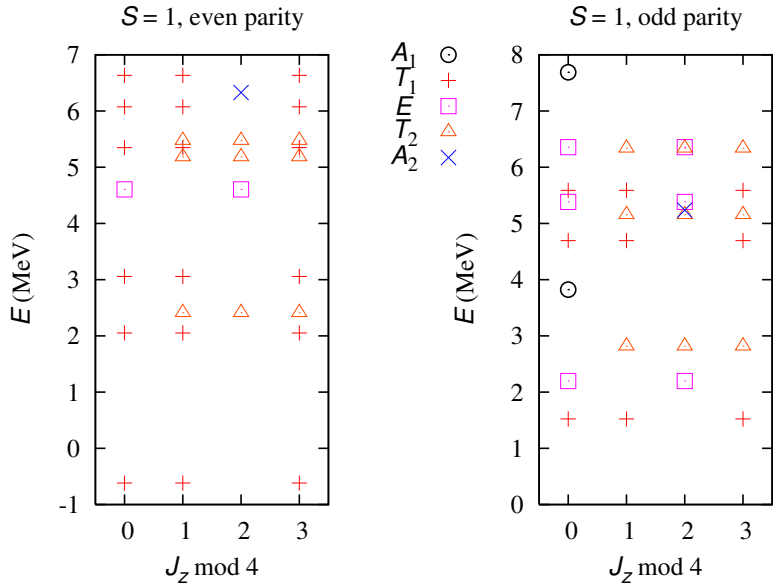}%
\caption{Interacting spectrum for $S=1$ standing waves for a $12\times
12\times12$ periodic lattice.}%
\label{s1_cube12}%
\end{center}
\end{figure}

The hard spherical wall removes extra copies of the two-particle interactions
due to the periodic boundaries of the lattice. \ This appears to be important
for measuring phase shifts and mixing angles at higher energies and higher
angular momentum. \ For comparison we show in Fig. \ref{s1_cube12} the same
test potential spectrum for intrinsic spin $S=1$ on a $12\times12\times12$
lattice with periodic boundaries.%
If we compare with Fig. \ref{s1_r10} we see that the SO$(3,Z)$ representations
in the various total angular momentum $J$ multiplets have been split apart
much further. \ It is easy to identify the lowest-lying standing waves but
difficult to clearly discern much beyond this.

The method we have presented can be applied directly to any nonrelativistic
effective theory of point particles on the lattice. \ There is a long list of
interesting few- and many-body systems one could consider on the lattice.
\ For example one interesting system is that of atoms and molecules with
long-range dipole interactions,%
\begin{equation}
V_{\text{dipole-dipole}}(\vec{r}\,)\varpropto-\frac{1}{4\pi r^{3}}\left[
3\left(  \hat{r}\cdot\vec{\mu}_{1}\right)  \left(  \hat{r}\cdot\vec{\mu}%
_{2}\right)  -\vec{\mu}_{1}\cdot\vec{\mu}_{2}\right]  .
\end{equation}
There is interest both in magnetic dipole interactions of cold atoms
\cite{Regal:2003B, Zhang:2004A, Schunck:2005, Gaebler:2007A,Harris:2007}\ and
electric dipole interactions in cold polar molecules \cite{Doyle:2004A,
Gunter:2005A, Micheli:2006A, Brennen:2007A}. \ These spin-changing
interactions are important for determining topological structures of the
many-body ground state as well as practical issues impacting evaporative
cooling in magnetic traps.

We will discuss in some detail the application to low-energy nuclear physics
in a forthcoming paper. \ One important low-energy interaction between
nucleons is the exchange of a virtual pion, producing spin-dependent forces of
the form%
\begin{align}
V_{1\pi}(\vec{r}\,)  &  =\left(  \frac{g_{A}}{2f_{\pi}}\right)  ^{2}\left(
\boldsymbol\tau_{1}\cdot\boldsymbol\tau_{2}\right) \nonumber\\
&  \times\left\{  \frac{m_{\pi}^{2}e^{-m_{\pi}r}}{12\pi r}\left[  S_{12}%
(\hat{r})\left(  1+\frac{3}{m_{\pi}r}+\frac{3}{(m_{\pi}r)^{2}}\right)
+\vec{\sigma}_{1}\cdot\vec{\sigma}_{2}\right]  -\frac{1}{3}\vec{\sigma}%
_{1}\cdot\vec{\sigma}_{2}\delta^{3}(\vec{r}\,)\right\}  .
\end{align}
Here $\boldsymbol\tau$ are Pauli matrices in isospin space, $m_{\pi}$ is the
pion mass, $f_{\pi}$ is the pion decay constant, $g_{A}$ is the nucleon axial
charge, and $S_{12}(\hat{r})$ is the tensor operator for spin-1/2 particles
defined in (\ref{s12}). \ Recent reviews of literature relating to chiral
effective field theory can be found in \cite{Bedaque:2002mn, Epelbaum:2005pn}.
\ This tensor interaction has some general similarities to that of the test
potential considered here. \ However the divergent short-distance behavior of
the one-pion exchange potential and other short-distance nucleon-nucleon
interactions means that the lattice spacing plays an essential role in
regulating ultraviolet divergences. \ The hard spherical wall method we have
presented should therefore be useful in determining the underlying physics of
a given lattice action. \ The singular short-distance interactions are likely
to cause larger breaking of rotational invariance than that found for the
bounded test potential. \ In such cases it may be useful to include
higher-order derivative interactions specifically designed to cancel lattice artifacts.

\section*{Acknowledgements}

Partial financial support from the Deutsche Forschungsgemeinschaft (SFB/TR
16), Helmholtz Association (contract number VH-NG-222), and U.S. Department of
Energy (DE-FG02-03ER41260) are gratefully acknowledged. \ This research is
part of the EU Integrated Infrastructure Initiative in Hadron Physics under
contract number RII3-CT-2004-506078.

\bibliographystyle{apsrev}
\bibliography{NuclearMatter}

\end{document}